\documentclass[sigconf,nonacm=True]{acmart}

\usepackage{tikz}
\usepackage{amsmath}

\usepackage{booktabs}
\usepackage{paralist}
\usepackage{subcaption}
\usepackage[group-separator={,}, per-mode=symbol]{siunitx}
\usepackage{color}
\usepackage{xspace}
\usepackage{eurosym}
\usepackage{graphicx}
\usepackage{color, colortbl}
\usepackage{amsfonts}
\usepackage{threeparttable, tablefootnote}
\usepackage{balance}

\newcommand{\ie}{i.\@\,e.\@\xspace}
\newcommand{\eg}{e.\@\,g.\@\xspace}
\newcommand{\etal}{et~al.\@\xspace}

\newcommand{\Paragraph}[1]{\vspace{1pt}\noindent{\bf #1.}}

\newcommand{\tline}[2][c]{%
\begin{tabular}[#1]{@{}l@{}}#2\end{tabular}}
\definecolor{Gray}{gray}{0.5}

\PassOptionsToPackage{hyphens}{url}\usepackage{hyperref}

\AtBeginDocument{%
  \providecommand\BibTeX{{%
    \normalfont B\kern-0.5em{\scshape i\kern-0.25em b}\kern-0.8em\TeX}}}

\copyrightyear{2022}
\acmYear{2022}
\setcopyright{rightsretained}
\acmConference[WiSec '22] {Proceedings of the 15th ACM Conference on Security and Privacy in Wireless and Mobile Networks}{May 16--19, 2022}{San Antonio, TX, USA.}
\acmBooktitle{Proceedings of the 15th ACM Conference on Security and Privacy in Wireless and Mobile Networks (WiSec '22), May 16--19, 2022, San Antonio, TX, USA}
\acmPrice{15.00}
\acmISBN{978-1-4503-9216-7/22/05} 

\acmDOI{10.1145/3507657.3528536}

\author{Paul Staat}
\affiliation{%
  \institution{Max Planck Institute for Security and Privacy}
  \city{Bochum}
  \country{Germany}}
\email{paul.staat@mpi-sp.org}

\author{Kai Jansen}
\affiliation{%
  \institution{PHYSEC GmbH}
  \city{Bochum}
  \country{Germany}}
\email{kai.jansen@physec.de}

\author{Christian Zenger}
\affiliation{%
  \institution{PHYSEC GmbH}
  \city{Bochum}
  \country{Germany}}
\email{christian.zenger@physec.de}

\author{Harald Elders-Boll}
\affiliation{%
  \institution{Technische Hochschule Köln}
  \city{Cologne}
  \country{Germany}}
\email{harald.elders-boll@th-koeln.de}

\author{Christof Paar}
\affiliation{%
  \institution{Max Planck Institute for Security and Privacy}
  \city{Bochum}
  \country{Germany}}
\email{christof.paar@mpi-sp.org}

\begin{document}
\fancyhead{}

\title{Analog Physical-Layer Relay Attacks with Application to Bluetooth and Phase-Based Ranging}

\begin{abstract}
Today, we use smartphones as multi-purpose devices that communicate with their environment to implement context-aware services, including asset tracking, indoor localization, contact tracing, or access control. As a de-facto standard, Bluetooth is available in virtually every smartphone to provide short-range wireless communication. Importantly, many Bluetooth-driven applications such as Phone as a Key~(PaaK) for vehicles and buildings require proximity of legitimate devices, which must be protected against unauthorized access. In earlier access control systems, attackers were able to violate proximity-verification through relay station attacks. However, the vulnerability of Bluetooth against such attacks was yet unclear as existing relay attack strategies are not applicable or can be defeated through wireless distance measurement. 

In this paper, we design and implement an analog physical-layer relay attack based on low-cost off-the-shelf radio hardware to simultaneously increase the wireless communication range and manipulate distance measurements. Using our setup, we successfully demonstrate relay attacks against Bluetooth-based access control of a car and a smart lock. Further, we show that our attack can arbitrarily manipulate Multi-Carrier Phase-based Ranging~(MCPR) while relaying signals over~\SI{90}{\m}.
\end{abstract}

\maketitle

\section{Introduction}

Today, billions of smart wireless devices, including the Internet of Things~(IoT), surround us in our daily live. Increasingly, the wireless environment is used beyond mere communication to provide context-aware services: Wireless devices are integrated into the environment and applications such as asset tracking, indoor localization, contact tracing, and access control bring physical aspects into the digital world. Here, devices must be aware of their environment and distance to others.

Not surprisingly and in line with the ongoing technological convergence, the smartphone is a central aspect of many context-aware applications. Using a generic platform eliminates the need of dedicated hardware devices while further enhancing convenience. For Phone as a Key~(PaaK)~\cite{ieeespectrum.2019, Ho.2016}, the smartphone acts as a personal token for access control systems of, \eg, vehicles and buildings. As a prominent example, during the Covid-19 pandemic, smartphone-enabled contact tracing applications have emerged worldwide to track infection chains based on the proximity of individuals~\cite{baumgartner_mindthegap_2020}. Such applications naturally have to make use of the wireless communication systems implemented in smartphones. Being available in virtually every smartphone, Bluetooth has become the technology of choice for the required ad hoc short-range networks. Within this trend, vendors now increasingly also utilize Bluetooth for security-critical proximity verification applications, \eg, access control. However, Bluetooth was originally designed for wireless communication rather than for secure proximity verification. Thus, one may suspect a vulnerability against \textit{relay attacks}.

In the classical relay attack, an attacker makes distant victim parties being able to communicate by establishing an artificial communication channel. The distant parties thus may falsely assume to be in each other’s proximity. In the past, this has been exploited to circumvent earlier-generation automotive Passive Keyless Entry~(PKE) systems~\cite{Francillon.2010}. Thus, attackers are able to gain access to vehicles and media frequently reports car thefts due to relay attacks~\cite{Bilton.2015, Greenberg.2016}. In contrast, relay attacks on Bluetooth-based proximity verification have received only very little attention as of yet. While the attack has been identified as a threat for smart locks~\cite{Ho.2016}, realizations are still limited to unidirectional forwarding of contact tracing advertisement packets~\cite{baumgartner_mindthegap_2020}. Other than this, full-blown bidirectional range-extending relay attacks -- required for challenge-response authentication -- have not been reported. Since Bluetooth uses a very different wireless physical-layer architecture than traditional automotive PKE systems, previous attack strategies~\cite{Francillon.2010} cannot directly be applied. Consequently, the lacking attack realization currently hampers risk assessment of Bluetooth w.r.t. practical relay attacks and leaves the exploration of threat potential as an open research problem.

To close this gap, in this work we identify technical peculiarities of relay attacks on Bluetooth and outline why previous attack strategies are not applicable. Based on our analysis, we utilize commodity Radio Frequency~(RF) components to design and implement an analog attacker setup to achieve bidirectional physical-layer relaying of time-division duplex~(TDD) wireless communication such as Bluetooth. Our cable-based proof-of-concept allows hands-on testing of products and enables real-world relay attacks in certain scenarios. We use our setup to test recent Bluetooth-based PKE systems of a car and a smart lock, both of which utilize smartphone-based key replacements. Our attack allowed us to unlock the door and start the engine (in case of the car) while the legitimate smartphone was at a distance of more than \SI{65}{\m}. Our results confirm that Bluetooth alone is vulnerable against simple range-extending relay attacks. 

As per its latest v5.3~specification~\cite{BluetoothSIG.January2019}, the only way to infer the distance between two Bluetooth devices is based on signal strength which is known to be notoriously inaccurate~\cite{giovanelliRSSITimeofflightBluetooth2018}. Therefore, vendors increasingly implement Multi-Carrier Phased-Based Ranging~(MCPR) aside of Bluetooth~\cite{AbidinSecureAccuratePractical_2021, Stitt.2020, Dialog.ranging, imec.undated, Zand.2019, denso.2020, alpsalpine.2021}, recently even within integrated circuits geared towards MCPR~\cite{Bechthum.2020, imec.rangingIC, synaptics.comboIC}. Since MCPR in the past has been marketed as High Accuracy Distance Measurement (HADM)~\cite{BluetoothSIG_HADM_MCPR.2021, everything_imec_HADM_MCPR.2018, denso.2020}, it is also interesting to note that the Bluetooth~SIG lists a specification under development bearing this very name~\cite{BluetoothSIG_HADM.2021}. In view of these developments, we additionally study the security of MCPR against our attacker setup, showing that MCPR is a viable detection method for simple range-extension attacks. However, adding a simple phase-shifter to our setup, we demonstrate a novel distance manipulation attack against MCPR. Thus, for the first time, we achieve both range-extension (bridging a physical distance of~\SI{90}{\m}) and distance manipulation \textit{at the same time}. Our results confirm previous findings of Ólafsdóttir~\etal~\cite{Olafsdottir.2017}, who first described vulnerabilities of MCPR against signal manipulation attacks. Further, we give guidelines to enhance attack difficulty, \eg, by leveraging frequency hopping or analyzing channel reciprocity.

\Paragraph{Differentiation from previous work}
The general idea for relay and distance manipulation attacks is not original to this paper. The vulnerability of access control against range-extending relay attacks was first demonstrated by Francillon~\etal~\cite{Francillon.2010} for earlier PKE systems, using unidirectional signal amplification. However, this strategy is insufficient to achieve the bidirectional range extension of TDD wireless communication such as Bluetooth, required due to challenge-response authentication. In our work, we realize a bidirectional amplification approach using power detection to toggle signal directions. Ólafsdóttir~\etal~\cite{Olafsdottir.2017} presented the first security analysis of MCPR, revealing vulnerabilities against signal manipulation. The authors proposed the general idea of manipulating the phase of individual MCPR tones, although not presenting a practical attack realization. The authors gave a brief technical proposal based on mixing with locally generated RF signals. We believe that this approach is of low practicality as an attacker needs to generate a coherent RF carrier before phase manipulation can be applied. Ólafsdóttir~\etal also acknowledge this challenge in their paper as they sketch a countermeasure based on randomized phase shifts which later was adopted by Abidin~\etal~\cite{AbidinSecureAccuratePractical_2021}. We build on the idea of manipulating individual MCPR tones~\cite{Olafsdottir.2017}, albeit using an entirely different signal manipulation mechanism based on a simple phase shifter circuit, eliminating the unrealistic requirement of a coherent RF carrier. Moreover, we contribute the method to properly schedule phase shifts. Not only does this facilitate a real-world implementation (as our results show), it also bypasses the previously suggested countermeasure. To the best of our knowledge, this is the first work to unify range extension~\cite{Francillon.2010, Hancke.2005b, Francis.2011, Roland.2012} and distance manipulation~\cite{Olafsdottir.2017, Ranganathan.2012, Flury.2010} -- previously addressed separately -- within a real-world implementation.

\Paragraph{Contribution} In summary, this paper makes the following contributions:
\begin{compactitem}
    \item We design a practical physical-layer relay attack, generically applicable to standard \SI{2.4}{GHz} TDD communication systems such as Bluetooth. We use a novel attack strategy which adapts to the transmit-receive timing of the legitimate parties. We present a prototypical cable-based implementation built from commodity RF components.
    
    \item We analyze the Bluetooth-based access control implementations of a car and a smart lock and demonstrate successful relay attacks against both systems. To the best of our knowledge, this is the first work to fully implement a relay attack to demonstrate the insecurity of Bluetooth-based access control.

    \item We investigate MCPR as a countermeasure against our attack and demonstrate a novel phase manipulation attack to simultaneously perform range extension and distance manipulation. %
\end{compactitem}

\Paragraph{Responsible Disclosure} We provided this paper to the Bluetooth SIG and the affected manufacturers.

\section{Related Work and Background}
In this section, we outline related works and provide background information on RF proximity verification, relay attacks, and Bluetooth communications. %

 \subsection{Related Work}
\label{sec:related}

The literature describes various relay attacks against different RF proximity verification systems. Francillon~\etal~\cite{Francillon.2010} have demonstrated signal amplification attacks to increase the range of immobilizing signals of classical automotive PKE systems. Their study revealed vulnerabilities on all tested models. Payment systems are subject to relay attacks as well and have been demonstrated for ISO~14443~\cite{Hancke.2005b} and NFC~\cite{Francis.2011, Roland.2012}. Relay attacks against Bluetooth communications have received little attention as of yet, although Levi~\etal~\cite{Levi.2004} discussed the possibility of such attacks already in 2004. In a security analysis of Bluetooth-based smart locks, Ho~\etal~\cite{Ho.2016} identified the attack as a threat, albeit not addressing technical realization. Baumgärtner~\etal~\cite{baumgartner_mindthegap_2020} demonstrated Software-Defined Radio~(SDR)-based unidirectional forwarding of advertisement packets to study the security of contact tracing applications. Protocol-level impersonation attacks as described by Antonioli~\etal~\cite{Antonioli.2020} and Jasek~\cite{SawomirJasek.} are conceptually different from a relay attack but likewise allow attackers to forward Bluetooth traffic between victim parties. 

To defeat relay attacks, numerous works study wireless distance measurements~\cite{Bensky.2016} and their security guarantees. Ranganathan and \v{C}apkun~\cite{Ranganathan.2017} survey techniques for secure ranging and conclude with Ultra Wideband Impulse Radio~(\mbox{UWB-IR}) as the most promising candidate. The fine time resolution of wideband waveforms promotes UWB systems to be used for Time-of-Flight~(ToF)-based ranging within distance-bounding protocols. While ToF itself cannot be reduced by an attacker, the measurement procedure can still be vulnerable: Clulow~\etal~\cite{Clulow.2006} have introduced the Early-Detect, Late-Commit~(ED/LC) attack that has later been applied to UWB-based ranging~\cite{Flury.2010}. Singh~\etal~\cite{Singh.2019a} and Leu~\etal~\cite{Leu.2019} recently introduced novel physical-layer security primitives to diminish the attacker's success while preserving the ability of long-range communication. Another work of Singh~\etal~\cite{Singh.2019b} aims to detect UWB distance-enlargement attacks. %
Apart from UWB, previous work also investigates the security of other RF proximity verification techniques. Ólafsdóttir~\etal~\cite{Olafsdottir.2017} presented the first security analysis of MCPR and demonstrated a delay-based distance reduction attack. In order to counteract these attacks, Abidin~\etal~\cite{AbidinSecureAccuratePractical_2021} recently combined MCPR with a coarse ToF measurement. Notably, they report a promising implementation on a standard Bluetooth transceiver. Ranganathan~\etal~\cite{Ranganathan.2012} have shown ED/LC attacks on chirp-based ToF measurements. 

Other than distance measurement, proposals for relay attack detection are RF fingerprinting~\cite{Joo.2020}, channel reciprocity~\cite{Jain.2012b, Krentz.2014}, protocol timing~\cite{Reid.2007}, and sensor fusion~\cite{Truong.2014, shrestha2014drone}. In 1993, Brands and Chaum~\cite{Brands.1993} introduced distance bounding protocols to cryptographically verify an upper bound on the distance of a prover. Finally, the literature reports additional vulnerabilities in access control systems. Eisenbarth~\etal~\cite{Eisenbarth.2008} demonstrated a side-channel based extraction of group keys from the KeeLoq system. Wouters~\etal point out security weaknesses in the implementations of access control systems of luxury cars~\cite{Wouters.2019, Wouters.2021}.

\subsection{Background}
\label{sec:background}

\Paragraph{Relay Attacks} In a classic relay attack, an adversary establishes a communication channel between distant parties that otherwise would not be able to communicate. Wireless radio systems are particularly prone to such attacks as radio wave propagation relies on a shared medium that the attacker can access, \ie, to capture/eavesdrop legitimate signals. To forward legitimate signals, attackers employ \textit{amplify-and-forward} or \textit{decode-and-forward} relaying schemes~\cite{Goldsmith.2005}. The latter involves recovery of bits or symbols from analog waveforms which is attractive for long-range transfers. In contrast, amplify-and-forward omits demodulation and forwards the analog waveform with minimal processing efforts, \eg, amplification only, resulting in low delays and moderate design complexity at the cost of link budget constraints. With the advent of distance measurement techniques, the scope of relay attacks became broader: The attacker is required to alternatively or additionally perform some sort of signal manipulation to overcome the implemented proximity verification. In turn, a location-based wireless service may falsely allow access to restricted applications, potentially causing economical damage to individuals and businesses. Examples for threatened services are wireless payment systems, contact tracing, electronic door locks, and PKE systems for cars.

\Paragraph{RF Proximity Verification} Wireless communication systems can be used for localization and ranging~\cite{Bensky.2016, Boukerche.2007, Zand.2019}. Leveraging physical-layer observations such as ToF, Received Signal Strength~(RSS), or carrier phase, devices infer distances to others. Distance measurements from ToF are based on the signal propagation delay being a function of the speed of light and distance. Accurate measurement of the ToF requires nanosecond time resolution as provided, \eg, by UWB-IR~\cite{Yassin.2016}. RSS-based ranging leverages the propagation path loss of radio waves as a function of the distance~\cite{giovanelliRSSITimeofflightBluetooth2018}. RSS indication is available for almost every wireless receiver but suffers from inaccuracies due to multipath propagation. Carrier phase-based ranging, \ie, MCPR, uses unmodulated RF carrier signals to observe phase shifts that are proportional to the ToF~\cite{Bensky.2016}. The measurement procedure can conveniently be implemented with narrowband frequency hopping systems such as Bluetooth Low Energy~(BLE)~\cite{Zand.2019} and is capable of accuracies below \SI{30}{cm}~\cite{Bechthum.2020}.

\Paragraph{Bluetooth} Bluetooth is designed for short-range wireless communication. The specification~\cite{BluetoothSIG.January2019} defines two wireless stacks, namely BLE and Bluetooth BR/EDR. In the following, we use the terms BLE and Bluetooth to refer to BLE devices. Still, our results also apply to BR/EDR in principle. BLE operates on a total of $40$~sub channels with~\SI{2}{\MHz} spacing within the~\SI{2.4}{\GHz}~ISM band and uses TDD to realize bidirectional communication. On the physical layer, transmissions use Gaussian Frequency Shift Keying (GFSK) with data rates of either \SI{1}{Mbps} or \SI{2}{Mbps}. Furthermore, BLE employs an adaptive frequency hopping scheme over $37$ channels with a hopping rate of up to approx.~\SI{133}{\Hz}. Device discovery in BLE is realized through periodic advertisement packets. In a subsequent \textit{pairing} phase, parties can establish an authenticated and encrypted channel, \ie, cryptographic keys are exchanged. The secure channel is persistent as both nodes store long term keys, which is referred to as \textit{bonding} of two devices in BLE. Bluetooth does not specify a dedicated measurement procedure to find the distance between two devices. Instead, RSS values can to be used to coarsely estimate distances. When multiple devices are available, an alternative approach is triangulation based on Angle-of-Arrival~(AoA) or Angle-of-Departure~(AoD) which was added to the Bluetooth specification v5.1. Both features leverage the Constant Tone Extension~(CTE) which was already used to implement MCPR~\cite{Zand.2019}.

\section{Relay Implementation}
\label{sec:relaying_bluetooth}

Next, we introduce the system and attacker model and outline challenges associated with relay attacks on Bluetooth. Finally, we outline our proof-of-concept analog physical-layer relay attack.

\subsection{System and Adversary Model}
We consider an external relay attacker operating on the wireless physical layer with the goal of circumventing Bluetooth-based RF proximity verification between a pair of distant communication parties~$\mathsf{A}$ and~$\mathsf{B}$. Considering an access control application, $\mathsf{B}$ is a mobile device, \eg, a smartphone, whereas~$\mathsf{A}$ is mostly stationary, \eg, a smart lock or a car. We assume that~$\mathsf{A}$ and~$\mathsf{B}$ are honest and have created a bond before, \ie, the attacker is not able to break the applied cryptography, that is to read or make valid manipulations of secured payload data. The nodes employ standard-compliant BLE communication and $\mathsf{A}$ needs to infer proximity to $\mathsf{B}$ from the ability to communicate and reasonable RSS levels. The attacker is capable of transmitting and receiving RF signals to and from both parties. Also, the attacker can choose a strategic position close to~$\mathsf{A}$. 

\subsection{Previous Attack Strategies}
\label{sec:analog_design_consideration}

Previous strategies for practical attacks were based on unidirectional relaying. While this is sufficient for inherently unidirectional systems, \eg, advertisement-based contact tracing~\cite{baumgartner_mindthegap_2020}, system designers can enforce bidirectional communication using challenge-response authentication. However, even then, unidirectional attacks can be sufficient, as is the case for earlier generation automotive PKE~\cite{Francillon.2010}: Here, a dedicated wireless protocol is used between a car and a key fob. The car transmits Low Frequency~(LF) signals at around \SI{100}{\kHz} to the key fob which responds on an Ultra-High Frequency~(UHF) channel at \SI{315} or \SI{433}{\MHz}. The LF signals only have short range and therefore \textit{immobilize} the system. In contrast, the UHF signals from the key have a wide range. The attack outlined by Francillon~\etal~\cite{Francillon.2010} exploits this observation, extending the LF signal range to reach the distant key. This simplifies the attack to a unidirectional amplification of the LF signals while the response from the key fob reaches the car directly. Importantly, the distinct signal frequencies allow attackers to separately manipulate each communication direction.

However, the situation is different with BLE-based communication. Here, the legitimate parties share the same frequency resource over time using TDD to establish bidirectional communication, having the same range in both directions. Thus, in order to extend the victim's effective communication range, a relay attacker faces the challenge of extending \textit{both} directions. Now, one may suggest to use a pair of BLE receivers and transmitters at both ends to accomplish this task by forwarding only the application payload data, \ie, decode-and-forward relaying. Unfortunately, this cannot be done since resource allocation of the legitimate parties, \eg, the transmit-receive timing and frequency hopping sequence\footnote{The BLE frequency hopping is randomized and synchronization is cumbersome due to the prediction of PRNG values~\cite{Cauquil.2019}.}, is unknown to the attacker who is completely external and cannot impersonate the legitimate parties. Further, decode-and-forward relaying neglects physical-layer information and is easily detected by, \eg, MCPR or channel reciprocity (see Section~\ref{sec:attack_evaluation}).

Alternatively, following the style of attack of~\cite{Francillon.2010} for both directions, the attacker could try to amplify and forward the entire Bluetooth spectrum at both legitimate parties towards the other. However, this approach requires careful design consideration since both communication directions use the same frequency range. That is, an attacker naively following this approach could end up building a large feedback loop that amplifies itself, eventually leading to instability and self-destruction.

\subsection{Relay Implementation}
\label{sec:relay_implementation}

\begin{figure}
\centering
\includegraphics[width=0.97\linewidth]{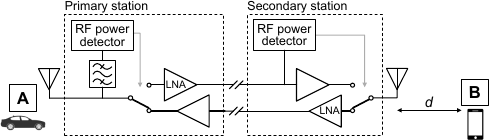}%
\caption{This block diagram illustrates the hardware setup for the analog relay attack, comprising of the two distant primary and secondary stations.}
 \label{fig:relay_block_only_PD}
\end{figure}

The attacker's goal is to ($i$)~establish a bidirectional communication channel while ($ii$)~applying amplification in both directions for range-extension and to circumvent RSS-based proximity verification. However, as outlined before, the relay attacker cannot simply amplify the legitimate BLE signals in both directions independently. To resolve this issue, we propose a novel relaying strategy which adapts to the legitimate node's transmit-receive behavior. An implementation block diagram is shown in Fig.~\ref{fig:relay_block_only_PD}. The attacker positions their antenna close to~$\mathsf{A}$ and uses an analog RF power detector to sense when~$\mathsf{A}$ transmits. Upon detection of a transmission of~$\mathsf{A}$, an RF switch is used to select the upper signal path, applying amplification and forwarding signals to~$\mathsf{B}$. In the other case that no transmission is detected, the lower signal path is selected, applying amplification in the other direction to forward signals from~$\mathsf{B}$ to~$\mathsf{A}$. This strategy allows the attacker to synchronize to the node's TDD, preventing simultaneous amplification in both directions to eliminate the aforementioned feedback loop.

The output of the power detector is a voltage proportional to the RF input power. Using a comparator, we implement a threshold-based binary power detection to sense transmissions of~$\mathsf{A}$. The comparator output logic signal controls an RF switch to change signal directions. Allowing the attacker to bridge relatively large distances between the two relay end-points, we apply this technique at both the primary and secondary relay stations, cf.~Fig.~\ref{fig:relay_block_only_PD}. At the primary side, close to~$\mathsf{A}$, the power detection takes place after the antenna but before the RF switch and signals of~$\mathsf{A}$ will reach the power detector regardless of the switch position. At the secondary side, power detection is again used to detect forwarded transmissions of~$\mathsf{A}$. Due to the attacker's close position to~$\mathsf{A}$, the signals of~$\mathsf{A}$ reach the primary power detector rather strongly. In contrast, after over-the-air and cable losses, the signals of~$\mathsf{B}$ arrive weaker and will not trigger the switching mechanism. Thus, the power-detection threshold should be selected below the expected high signal power from~$\mathsf{A}$ and above the low signal power from~$\mathsf{B}$.

Combining standard and ready-to-use RF components, we implemented a prototypical cable-based attacker setup that is depicted in Fig.~\ref{fig:analog_relay} in Appendix~\ref{sec:appendix}. We give an outline of implementation details and example traces of power detector output signals in Appendix~\ref{sec:appendix}. Further, the exact parts are listed in Table~\ref{tab:relay_part_list} likewise in the appendix. While we have not optimized the design for low cost, we estimate the total cost for the parts to be about~\EUR{2200}. However, omitting the evaluation boards, the individual components and ICs can be purchased for under~\EUR{1200}.

The maximum distance the attacker can bridge can be found by means of a link budget analysis, where the wireless loss to and from the relay antennas as well as the total amplification gain and loss of the relay hardware needs to be taken into account. Note that our attacker implementation is based on a mixed-signal design with the relay channel being completely analog. As no sampling is involved, the attacker operates with minimum latency which is desired for attacks on proximity verification. On the other hand, targeted and intelligent signal manipulation is more difficult, yet possible as we show in Section~\ref{sec:relay_phase_manipulation}. Another aspect related to the analog nature is the impact of interference. In our setup, \textit{all} signals picked up by the receiving antenna are forwarded to the other relay station, including signals from the legitimate and other parties. Thus, multiple wireless devices in reach of the relay stations won't diminish the attacker's success. Still, interference from other devices may affect the legitimate receiver, as is always the case in multi-user wireless settings. However, on the primary relay side, strong interfering signals could falsely trigger the power detector. We utilize a \SI{2.4}{\GHz} bandpass filter on the power detector input, rejecting interference from other frequency bands. To mitigate in-band interference, the attacker should ensure that the victim signals are the strongest received signals, \eg, by means of short distances or directional antennas. In our experiments, we did not encounter performance degradation due to interference, even though operating in busy wireless environments. For instance, close-by \SI{2.4}{\GHz}~\mbox{Wi-Fi} routers did not trigger the relay's power detection (cf.~Fig.~\ref{fig:advertising_interference} in the appendix). Finally, in view of PaaK systems, please note that Bluetooth and \mbox{Wi-Fi} do not transmit simultaneously~\cite{classen_wireless_coexistence}.

\section{Attacks on Bluetooth-Based Access Control}
\label{sec:attacks}
In this section, we test our attacker setup in two case studies on Bluetooth-based access control systems that both implement smartphone-based PKE. In particular, we analyzed the behavior of a car and an electronic door lock for buildings. Both products allow the smartphone to be used as a personal entry token, not requiring any user interaction for unlocking doors or starting the engine.

\subsection{Case Study 1: Car}
\label{sec:tesla}
For our first case study, we tested the Bluetooth-based PKE system of a car. The car accepts smartphones as an access tokens, allowing to unlock the doors and start the engine. The car uses Bluetooth to monitor the presence of an authorized smartphone. The user only needs to possess the phone and is not required to actively participate in the unlocking procedure.

Initially, we tested the system in a non-adversarial setting. With the authorized smartphone, we approached the locked car towards the passenger side until it unlocked. Then, we moved away from the car until it locked again. The observed distances are given in Table~\ref{tab:car_lock_unlock} and it is evident that different conditions apply for locking and unlocking. Since BLE is used, proximity verification is likely accomplished through analysis of RSS\footnote{AoA and AoD methods can also be utilized to find distances, however, requiring multiple locator devices to perform triangulation.}. Thus, we suspect that the car applies different RSS thresholds for unlocking and locking. The RSS hypothesis is further backed by the distances in Table~\ref{tab:car_lock_unlock} varying significantly with the phone location: The distances are largest when the phone is carried in the hand (strong line-of-sight channel) and lowest when the phone is carried in a trouser pocket on the back of the body (weak non-line-of-sight channel due to human body shadowing). Further, we observed that unlocking the car does not imply the ability to start the engine. In line-of-sight conditions, we were able to start the engine with the smartphone at a distance of \SI{2}{\m} from the car, hinting a third RSS threshold.

\begin{figure}
\centering
\hspace*{\fill}%
\subfloat[]{{%
        \includegraphics[width=0.47\columnwidth]{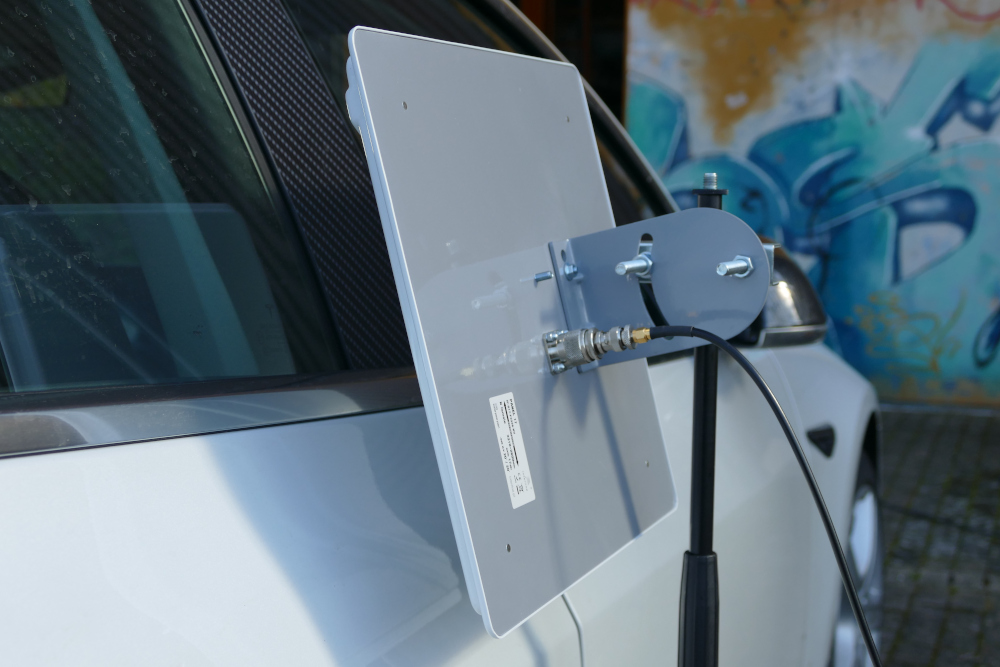}}}
\hfill
\subfloat[]{{%
        \includegraphics[width=0.47\columnwidth]{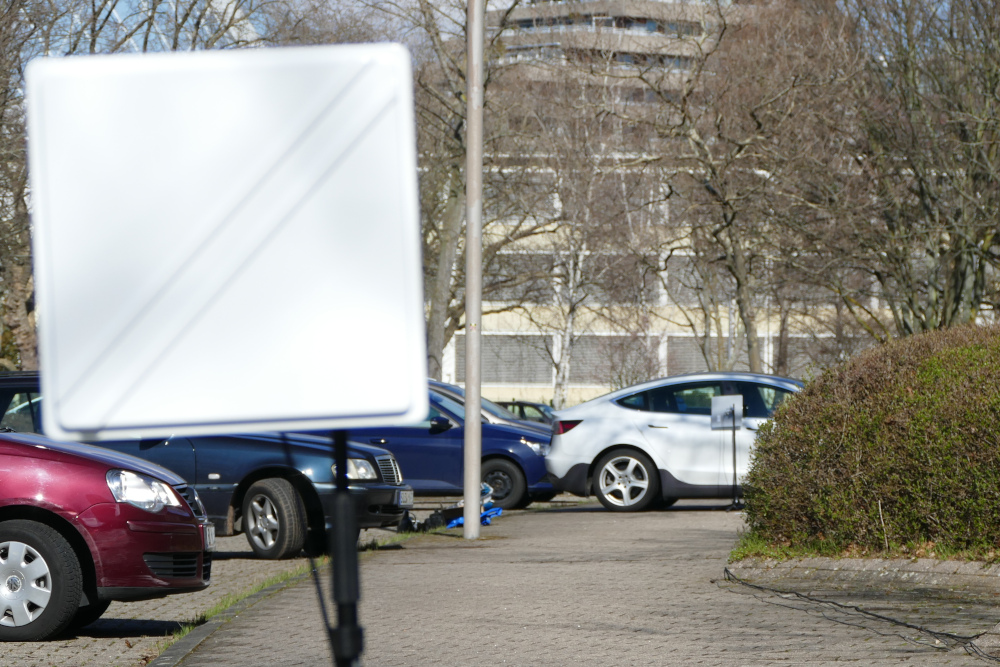}
        }}
\hspace*{\fill}%
\caption{Attack setup, relaying Bluetooth signals between the primary (a) and secondary (b) relay stations to establish communication between a car and a distant smartphone.}
\label{fig:tesla}
\end{figure}

\begin{table}
\footnotesize
\caption{Observed non-adversarial car unlocking and locking distances.}
\label{tab:car_lock_unlock}
\centering
\begin{tabular}{@{}rcc@{}}
\toprule
\textbf{Phone Location} & \textbf{Unlock} & \textbf{Lock}\\
\midrule
Hand                    & \SI{5}{m}       & \SI{13}{m}\\
Trouser pocket          & \SI{3}{m}       & \SI{11}{m}\\
Trouser pocket (back)   & \SI{1}{m}       & \SI{6}{m}\\
Jacket pocket           & \SI{4}{m}       & \SI{12}{m}\\
\bottomrule 
\end{tabular}
\end{table}

Next, we tested the effectiveness of our attacker setup outlined in Section~\ref{sec:relay_implementation}. We placed the primary relay station close (\SIrange{10}{30}{\cm}) to the B-pillar of the car. We placed the secondary relay station at a distance of approx.~\SI{65}{\m} to the car. Both relay stations were equipped with a directional antenna as can be seen from Fig.~\ref{fig:tesla}, showing the experimental attack setup. Approaching the secondary relay station antenna with the authorized smartphone in the hand, the car unlocked at a distance of \SI{4}{\m} to the relay antenna. Being at a distance of \SI{2}{\m} to the relay antenna, it was possible to start the engine. Also, we observed the previously outlined behavior for unlocking which took place when the smartphone was at a distance of \SI{13}{\m} to the relay antenna.

Our attack clearly succeeds to substantially extend the Bluetooth communication range between the car and the smartphone. We were able to unlock and start the car with the authorized smartphone being at a distance of \SI{69}{\m} and \SI{67}{\m}, respectively. Consequently, we circumvented the Bluetooth-based RF proximity verification, highlighting yet another attack vector to compromise keyless entry systems. During our experiments, we have not perceived any indication of additional countermeasures. Several approaches may be feasible, including GPS position comparison using the car's cellular network connection or plausibility checks of RSS values across the various Bluetooth antennas of the Model~3. However, as researchers have already stressed for a long time, using RSS for proximity verification is generally considered insecure.

We carried out all tests and attacks with a Tesla Model 3 Long Range from 2019, firmware version v10.2 (2021.4.12). The smartphone was a Huawei~P~Smart FIG-LX1 running Android~9.1.0.

\subsection{Case Study 2: Smart lock}

For our second case study, we tested a smart lock which is a Bluetooth-enabled keyless entry system which retrofits traditional key-based doors. %
It is controlled using a smartphone app and also supports an optional PKE mode which automatically unlocks the door upon proximity of an authorized smartphone. Similarly to the previous experiments with the car, we placed the primary relay station close to the smart lock. Then, with our setup we again bridged a distance of approx.~\SI{65}{\m} between an authorized smartphone and the smart lock. Now, with the smartphone as close as~\SI{2}{\m} to the secondary relay station, the smart lock unlocked the door. Again, our attacker setup successfully circumvents the Bluetooth-based RF proximity verification.

Compared to the previous car entry system, a smart lock has the distinct advantage of being deployed in a fixed location. This allows the smartphone to perform a number of plausibility checks before sending an unlock command to the lock. During our relay attack attempts, we noticed that the lock's smartphone app leverages multiple such checks. To analyze the behavior in detail, we took advantage of plain-text log files provided by the app, originally intended for diagnostic purposes. Notably, the log files in some cases even provide commentary insight to the application logic. From our analysis, we identified several measures and conditions when the lock is configured to automatically unlock: (1) Low Bluetooth transmission power is used and the application obtains RSS values. (2) The user must first exit and then re-enter a pre-defined geofence area before unlock. Distance to the lock is validated using GPS data and the known fixed device location. (3) The android app leverages Google Play services~\cite{google_play_services.2021} to monitor device activity such as walking. (4) Plausibility checks are applied, \ie, when location changes occur too fast or despite no movement being detected. Clearly, the outlined plausibility checks increase the hurdles for a successful attack. However, as GPS positions are considered insecure and may be spoofed, we conclude that attacks are still possible.

We carried out the tests and attacks with a Nuki Smart Lock~2.0, firmware version 2.9.10. As the authorized smartphone, we used a Huawei~P10~VTR-L09 running Android~9.1.0 with Nuki Smart Lock app version 2.7.4.

\section{Attacks On Phase-based Ranging}
\label{sec:attack_evaluation}
To counter attacks against Bluetooth-based proximity detection, RF-based physical distance measurement techniques can be employed. RSS-based distance estimates as per the current Bluetooth specification~\cite{BluetoothSIG.January2019} are inaccurate and prone to manipulations, \eg, by signal amplification like demonstrated in Section~\ref{sec:attacks}. Therefore, Bluetooth recently is increasingly being complemented by MCPR~\cite{AbidinSecureAccuratePractical_2021, Stitt.2020, Dialog.ranging, imec.undated, Zand.2019, denso.2020, alpsalpine.2021}, enabling distance measurements with high accuracy. In this section, we outline how our analog relay attack can be used to simultaneously increase the communication range while arbitrarily manipulating distances measured using MCPR.

\subsection{Background on MCPR}
MCPR is based on single carrier phase measurements, thus enabling accurate distance finding with narrowband radios. Notably, MCPR would be particularly attractive to implement with Bluetooth radios as the frequency hopping necessary for MCPR can be reused from or shared with the Bluetooth signaling, \eg, by using the CTE. %

\begin{figure}
  \centering
     \includegraphics[width=0.93\linewidth, trim={0.0cm 0.15cm 0cm 0cm},clip]{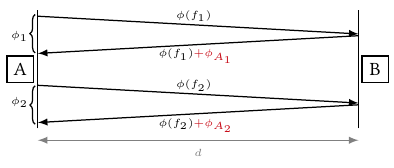}
\caption{Illustration of the MCPR measurement procedure between nodes~$\mathsf{A}$ and~$\mathsf{B}$ and potential phase manipulation attack using by applying adversarial phase shifts~$\phi_{A_1}$ and~$\phi_{A_2}$.}
  \label{fig:ranging_attack_illustr}
\end{figure}

In order to find the distance, the radios exchange unmodulated carrier signals and observe their phase shifts~\cite{Bensky.2016}. In particular, node~$\mathsf{A}$ transmits a tone to node~$\mathsf{B}$, which reflects the received tone back to~$\mathsf{A}$ (see Fig.~\ref{fig:ranging_attack_illustr}). Upon reception, $\mathsf{A}$ detects the phase of the response and can indirectly measure the ToF to calculate the distance to~$\mathsf{B}$. The idealized two-way tone exchange at frequency~$f_i$, assuming a line-of-sight channel, leads to a phase delay of~$\phi_i = 4\pi f_i \tau$ which is a function of twice the ToF~$\tau$ (and the distance between the nodes). As the carrier phase wraps with~$2\pi$, measurements at two frequencies~$f_1$ and~$f_2$ resolve distance ambiguities~\cite{Olafsdottir.2017} and allow~$\mathsf{A}$ to calculate its distance~to~$\mathsf{B}$:
\begin{equation}
    d = \frac{c_0}{4\pi} \frac{\phi_2 - \phi_1}{f_2 - f_1} = \frac{c_0}{4\pi} \frac{\Delta\phi}{f_\mathrm{step}} 
    \label{eq:ranging_distance}
\end{equation}
In practice, the measurement is repeated for multiple frequencies in rapid succession to combat noise and multipath distortion, \eg, by sweeping over $N$~frequencies to gather $N-1$~distance estimates for averaging. %

\subsection{Distance Manipulation}
\label{sec:mcpr_attack}
Ólafsdóttir~\etal~\cite{Olafsdottir.2017} have shown that MCPR has a number of vulnerabilities rooted in the feasibility to manipulate the signal phase. In turn, attackers may be able to manipulate MCPR distance measurements. One attack concept relies on individual phase manipulation of each carrier signal of a ranging procedure as illustrated in Fig.~\ref{fig:ranging_attack_illustr}. Here, we follow this general idea.

We observe from Eq.~\ref{eq:ranging_distance} that the distance measured by the nodes depends on the phase \textit{change} over frequency. Thus, an attacker attempting to manipulate the distance needs to forge a malicious phase slope by manipulating the channel phase response. By applying appropriate phase shifts~$\phi_{A_i}$, the legitimate parties will estimate the distance to be~$d_\mathrm{set}$:
\begin{equation}
    d_\mathrm{set} = \frac{c_0}{4 \pi} \frac{\Delta \phi + (\phi_{A_2} - \phi_{A_1})}{f_\mathrm{step}} = \frac{c_0}{4 \pi} \frac{\Delta \phi + \Delta \phi_{A}}{f_\mathrm{step}}
    \label{eq:ranging_distance_forged}
\end{equation}
Combining Eq.~\ref{eq:ranging_distance} and Eq.~\ref{eq:ranging_distance_forged}, we arrive at the following expression which yields the phase slope the attacker needs to apply to deliberately manipulate the distance measured by the legitimate parties to be~$d_\mathrm{set}$: 
\begin{equation}
    \Delta \phi_A = \frac{4\pi f_\mathrm{step}}{c_0} \left( d_\mathrm{set} - d \right)
    \label{eq:attacker_phase_changes}
\end{equation}
Thus, the attacker only needs to know the frequency step size $f_\mathrm{step}$ and the actual node distance $d$, which are either known or can be observed. Since $\Delta \phi_A$ is derived from Eq.~\ref{eq:ranging_distance}, it implicitly assumes a line-of-sight channel (having linear phase) between~$\mathsf{A}$ and~$\mathsf{B}$.

\subsubsection{Attack Implementation}
To realize the outlined attack, Ólafsdóttir~\etal~\cite{Olafsdottir.2017} proposed mixing of the victim signals with locally generated coherent RF carrier signals to realize the required phase shifting. However, this requires the attacker to first synchronize to the victim signals on the carrier phase level. If possible at all (note the \SI{}{\us} timing of MCPR tones), this approach is highly complex and would be costly and cumbersome to implement. Instead, we introduce a new attack variant which only requires a coarsely aligned timing instead of RF carrier synchronization, therefore allowing straightforward implementation.

\begin{figure}
\centering
\includegraphics[width=0.98\linewidth]{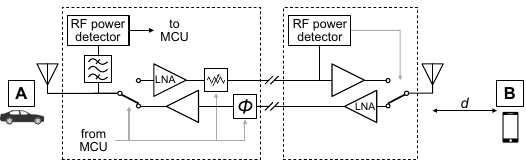}%
\caption{This block diagram illustrates the hardware setup for the analog relay attack with amplitude and phase manipulation capabilities.}
 \label{fig:relay_block_with_phase}
\end{figure}

We start with our attacker setup from Section~\ref{sec:relay_implementation} to which we add signal manipulation capabilities as shown in Fig.~\ref{fig:relay_block_with_phase}. We insert a digital phase shifter~\cite{MACOM.} into one path of the relay, adding phase control to signals traveling from~$\mathsf{B}$ to~$\mathsf{A}$. Please note that Fig.~\ref{fig:relay_block_with_phase} represents our experimental realization, with the attenuator and phase shifter in separate signal directions. However, insertion into the signal paths is possible in arbitrary other combinations as well.

To achieve distance manipulation, the phase shifter needs to be adjusted dynamically for each carrier signal, according to Eq.~\ref{eq:attacker_phase_changes} to forge a phase slope. To do so, we take advantage of the relay's power detector that we use to obtain the legitimate transmit-receive timing: As soon as $\mathsf{A}$ transmits, the rising power detector output indicates to the attacker that $\mathsf{B}$ is possibly about to respond. Thus, we can properly schedule the setting of phase shifts which need to change by~$\Delta \phi_A$ (see Eq.~\ref{eq:attacker_phase_changes}) each time the victim parties proceed with the next carrier frequency. Thereby, a coarse time synchronization to the tone exchange is established. Assuming a linearly increasing carrier frequency, \ie, \mbox{$f_i = f_{i-1} + f_\mathrm{step}$}, the phase shifter setting at time $t$, as dictated by the power detector output, is given by:
\begin{equation}
\label{eq:phase_shifter_settings}
    \phi_t = \phi_{t-1} + \Delta \phi_A\ \textrm{mod}\ 2\pi
\end{equation}
Applying this scheme in practice, the interested reader is referred to Appendix~\ref{sec:appdx_ranging_detector_output}, where we show exemplary output signals of the attacker's power detector during an MCPR procedure. 

\subsubsection{Attack Evaluation}
\label{sec:relay_phase_manipulation}
We now detail the attack performance in a practical evaluation. We conduct experiments with a commercially available MCPR implementation for BLE radio transceivers~\cite{Dialog.ranging} and demonstrate successful distance manipulation attacks while enhancing the communication range. The transceivers interleave standard BLE communication with an MCPR procedure with $N=40$~carriers and a frequency step size~$f_\mathrm{step}=$~\SI{1}{\MHz}.

\begin{figure}
  \centering
    \includegraphics[width=0.94\linewidth]{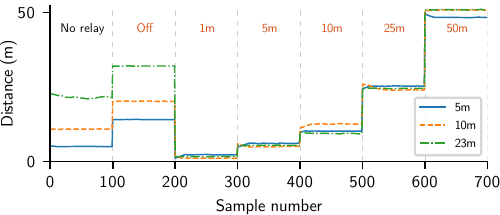}
  \caption{Demonstration of real-time distance manipulation, including reduction and enlargement for three victim node positions.}
  \label{fig:distacne_decreasing}
\end{figure}

In a first experiment, we study the general attack principle and focus on the effectiveness of adversarial signal manipulations. To ensure stable channel conditions and prevent direct radiation, we connect the RF port of the interrogating node~$\mathsf{A}$ directly to the primary side of the relay. We place the reflecting node~$\mathsf{B}$ in line-of-sight to the secondary relay antenna at distances $d$~of~\SI{5}{\m},~\SI{10}{\m}, and~\SI{23}{\m}. For each setting, we take~$100$ distance measurements. We plot the results in Fig.~\ref{fig:distacne_decreasing} with indication of the relay settings at the top. Without the relay, the distance measured by the legitimate nodes corresponds to the correct physical distance, as shown in the leftmost portion of Fig.~\ref{fig:distacne_decreasing}. Next, we insert the attacker hardware but disabled any distance manipulation. The measured distances (labeled as 'Off' in the plot) are now offset by a constant distance bias of around~\SI{9}{\m}. This offset is due to the relay's hardware time delay of around~\SI{30}{\ns} (approx.~\SI{23}{\ns} when excluding coaxial feed cables of~\SI{2}{\m} length). By estimating the physical node distance~$d$ accordingly higher, the attacker can easily self-compensate this effect as we will see next. 
For each node distance, we then configure the relay to manipulate the measured distances $d_\mathrm{set}$~to~\SI{1}{\m},~\SI{5}{\m},~\SI{10}{\m},~\SI{25}{\m}, and~\SI{50}{\m} by adjusting the phase shifter setting based on Eqs.~\ref{eq:attacker_phase_changes} and~\ref{eq:phase_shifter_settings}. As evident from Fig.~\ref{fig:distacne_decreasing}, the attacker succeeds to arbitrarily increase and decrease the measured distances accurately, regardless of the actual node distance. Please note that in our experiment, the attacker only was aware of~$d$ and~$f_\mathrm{step}$ but not of the actual MCPR results of~$\mathsf{A}$ and~$\mathsf{B}$, \ie, we did not adjust the phase shifter setting to match the desired distances. 

\begin{figure}
  \centering
    \includegraphics[width=0.97\linewidth]{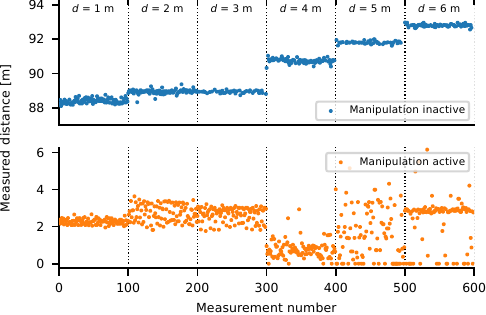}
  \caption{Combined relay and distance decreasing attack to enhance the communication range while simultaneously decreasing MCPR distance estimations of a BLE transceiver. Top: Measured distance without distance manipulation. Bottom: Measured distance with distance manipulation. %
}
  \label{fig:relay_ranging_attack_OTA}
\end{figure}

In the previous experiment, we investigated the attack principle and therefore granted the attacker ideal conditions, as node~$\mathsf{A}$ was directly connected to the relay through a coaxial cable. %
We now test the attack under more realistic conditions for the attacker. The attacker now picks up the signals from node~$\mathsf{A}$ wirelessly. We place the nodes~$\mathsf{A}$ and~$\mathsf{B}$ outside of their communication range in non-line of sight. In between the nodes, we install our analog relay attacker setup with off-the-shelf directional antennas to span a distance of~\SI{86}{m}. For the experiment, we place node~$\mathsf{A}$ at a fixed distance of~\SI{1}{\m} to the primary relay antenna. We place node~$\mathsf{B}$ at distances $d$~of~\SI{1}{\m} to~\SI{6}{\m} to the secondary relay antenna. Hence, the total distance between the nodes adds up to~\SIrange{88}{93}{\m}. Like in the attack demonstrations from Section~\ref{sec:attacks}, the nodes are only able to communicate via BLE due to the communication channel provided by the relay attacker. 
Next, the legitimate nodes take~$100$ MCPR distance measurements without adversarial distance manipulation. Fig.~\ref{fig:relay_ranging_attack_OTA}~(top) shows the results for the positions of node~$\mathsf{B}$~(indicated by labels at the top). As expected, the measured distance corresponds to the actual (relayed) distance between the nodes and rises as the distance to the relay antenna is increased. Thus, attacks like in Section~\ref{sec:attacks} would easily be detected. We now enable the distance manipulation with $d_\mathrm{set} =$~\SI{2}{\m}. The corresponding distance measurements in Fig.~\ref{fig:relay_ranging_attack_OTA}~(bottom) clearly indicate the attacker's success. This experiment highlights that our attacker implementation enables BLE communication over substantial distances while \textit{simultaneously} manipulating the MCPR procedure in real-time.

A key observation to make is that the sweeped carrier measurement for MCPR constitutes a particular security weakness: The channel transfer function is sampled on multiple frequencies consecutively which in turn allows individual manipulation of each carrier in a divide-and-conquer manner. This weakness becomes even more severe as, to the best of our knowledge, all currently available MCPR implementations utilize a non-randomly stepped RF carrier. In turn, simple attack strategies that do not require frequency knowledge, cf.~Eq.~\ref{eq:phase_shifter_settings}, can be applied. Therefore, we suggest that future MCPR deployments should use secure randomized frequency hopping sequences.

\subsection{Channel Reciprocity-based Detection}
\label{sec:channel_reciprocity_detection}
Besides carrier phase measurements, MCPR typically also provides the carrier amplitudes. While these are not necessary to infer the distance, it is still possible to evaluate them to perform a plausibility check based on \textit{channel reciprocity}. This fundamental property of radio wave propagation states that a radio channel between two antennas is symmetric~\cite{Balanis.2012}. Hence, signals sent on the same frequency from~$\mathsf{A}$ to~$\mathsf{B}$ and vice versa from~$\mathsf{B}$ to~$\mathsf{A}$ experience the same random propagation effects such as loss, phase shift, and multipath propagation, \eg, represented by a frequency-dependent complex-valued transfer function~$H(f)$.

Based on the claim that a relay attacker violates channel reciprocity between legitimate nodes~$\mathsf{A}$ and~$\mathsf{B}$, previous works have suggested to examine the channel response to detect relay attacks~\cite{Jain.2012b, Krentz.2014, Zenger.2016}. The attack detection mechanism is constructed from an examination of channel symmetry using a dissimilarity metric~$d$ on pairs of bidirectional channel magnitude responses $|H_{\textrm{AB}}(f)|$ and $|H_{\textrm{BA}}(f)|$. These need to be exchanged by the nodes via their authenticated and encrypted communication channel. A detection threshold~$\epsilon$ accounts for allowed differences in the respective channel observations made by~$\mathsf{A}$ and~$\mathsf{B}$, \eg, due to noise and device-dependent hardware imperfections: %
\begin{equation}
    d(|H_{\textrm{AB}}(f)|,\ |H_{\textrm{BA}}(f)|) > \epsilon
    \label{eq:cr_detection}
\end{equation}
Based on this condition, channel asymmetries introduced by an attacker can be detected. For instance, due to spatial decorrelation~\cite{Goldsmith.2005}, this is the case in \textit{unidirectional} attacks~\cite{Olafsdottir.2017, Ranganathan.2012}. Here, the nodes~$\mathsf{A}$ and~$\mathsf{B}$ are within their mutual communication range while the adversary intercepts (and manipulates) one communication direction while the other remains unchanged.

\begin{figure}
    \centering
    \includegraphics[width=0.95\linewidth]{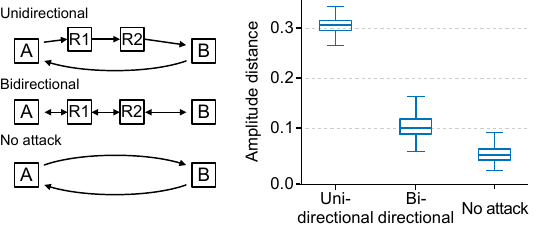}
    \caption{Left: Illustration of unidirectional and bidirectional relay attacks. Right: Distribution of Euclidean distances to assess channel reciprocity under unidirectional and bidirectional distance decreasing relay attacks.}
    \label{fig:attack_comparison}
\end{figure}

We put this claim to test in unidirectional and bidirectional (see left side of~Fig.~\ref{fig:attack_comparison}) distance decreasing attacks against MCPR between two BLE transceivers~\SI{15}{\m} apart from each other in an ordinary environment. First, we use the previously outlined attacker setup for a unidirectional relay attack. The attacker here only forwards signals from~$\mathsf{A}$ to~$\mathsf{B}$ while applying on-the-fly phase manipulation\footnote{We take the asymmetric channel into account to find the corresponding phase shifts.} to reduce the measured distance to~\SI{2}{\m}. Then, we repeat the attack but use bidirectional relaying where both communication directions are forwarded. Finally, we take a reference measurement without an attack with the nodes at an actual distance of~\SI{2}{\m}. We evaluate the channel reciprocity using the Euclidean distance between $|H_{\textrm{AB}}(f)|$ and $|H_{\textrm{BA}}(f)|$. Fig.~\ref{fig:attack_comparison}~(right) depicts the distributions of Euclidean distances for the three scenarios. As expected, channel reciprocity is violated most by the unidirectional attack. In contrast, the bidirectional attack exhibits increased channel reciprocity which overlaps with the results of the legitimate reference measurement.

The residual channel dissimilarity of our bidirectional attack is caused by differences between the relay's forward and reverse transmission paths, \eg, because of tolerances of the used parts. That is, in case of our attack, $|H_{\textrm{AB}}(f)|$ and $|H_{\textrm{BA}}(f)|$ now comprise of the concatenation of the reciprocal wireless channels to and from the relay antennas and the slightly non-reciprocal relay behavior. To completely circumvent attack detection, the relay hardware should be reciprocal, \ie, symmetric. This can be achieved in an additional engineering step by tuning the relay's transmission behavior to prevent Condition~\ref{eq:cr_detection} to be fulfilled. However, again shining light on the security drawbacks of the MCPR measurement principle, an attacker can also manipulate the measurement of $|H_{\textrm{AB}}(f)|$ and $|H_{\textrm{BA}}(f)|$ to diminish the effect of the imperfect relay hardware.

We propose a relay hardware equalization scheme where the attacker individually manipulates the amplitude of the $N$~tones exchanged by~$\mathsf{A}$ to~$\mathsf{B}$~(cf.~Fig.~\ref{fig:ranging_attack_illustr}). The approach follows a similar rationale as the previously outlined phase manipulation attack: In a divide-and-conquer manner, the attacker applies $N$~separate amplitude manipulations to each MCPR carrier. In this way, a frequency-dependent attenuation profile $\beta(f)$ can be applied, such that $|H_{\textrm{AB}}(f) \cdot \beta(f)| \approx |H_{\textrm{BA}}(f)|$ holds. Fortunately, $\beta(f)$ does not depend on the (random) wireless channels between the relay and the legitimate parties, as those are reciprocal either way. Instead, the attacker can determine $\beta(f)$ based on the the forward and reverse transmission behavior of the relay hardware, \eg, using a vector network analyzer. %
We apply $\beta(f)$ using a digital step attenuator that we insert into one path of the relay (see Fig.~\ref{fig:relay_block_with_phase}).

To identify the current tone frequency, we exploit the deterministic nature of the examined MCPR implementation: ($i$)~The MCPR measurement is always preceded by consistent transmit-receive patterns, allowing to identify the beginning of the tone exchange from analyzing the relay's power detector output signals. ($ii$)~The MCPR tone exchange is a linear frequency sweep with step size~$f_\mathrm{step}$, \ie, $f_i = f_{i-1} + f_\mathrm{step}$. Thus, the attacker can easily infer the current tone frequency from counting the number of transmissions.

\begin{figure}
    \centering
    \includegraphics[width=0.87\linewidth]{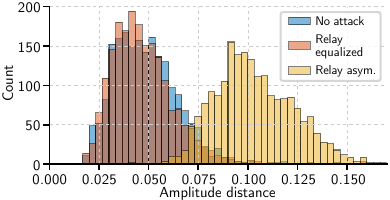}
    \caption{Distributions of Euclidean distances of channel responses, showing the effect of adversarial gain equalization in relation to non-equalized and non-adversarial settings.}
    \label{fig:relay_equalize}
\end{figure}

We repeat the MCPR procedure with and without the outlined amplitude manipulation attack as well as in a legitimate scenario. For all three cases, we plot the histograms of Euclidean distances of channel measurements in Fig.~\ref{fig:relay_equalize}. The distributions for the non-adversarial and the equalized relay settings are in complete agreement and are barely distinguishable. This demonstrates the effectiveness of the proposed equalization scheme to prevent attack detection from examining channel reciprocity. Overall, channel reciprocity can be helpful in thwarting naive attacks. However, we conclude that it only increases the attack difficulty, and thus, skilled attackers will be able to overcome this hurdle, as previously discussed and now demonstrated.

\section{Discussion}
\label{sec:discussion}
In this section, we discuss the real-world applicability, attacker capabilities and hardware improvements, and reason about attack detection and mitigation. Finally, we give directions for future work.

\subsection{Real-World Applicability}
A real-world relay attack is likely to be carried out against access control systems offering PKE operation as those do not require any user interaction other than proximity. However, manufacturers usually allow to disable this feature at will. Some products implement geofencing, which is insufficient since GPS positions are considered to be insecure and may be spoofed. Given the availability of low-cost, credit-card sized software-defined radios~\cite{hackrf_one} and ready-to-use software~\cite{gpsspoofing.git}, we believe that such an attack could be carried out in parallel to relaying with little effort.

While the cable-based relay implementation certainly rules out some attack scenarios, this should pose a modest hurdle for willing and prudent attackers, \eg, when attacks take place at night times. Owing to the operation principle, our relay requires input signals strong enough for the power detector-based reactive switching. Thus, the primary relay station needs to be in proximity to one victim node. This is a realistic assumption, as proximity prior to authorization is an integral part of access control. In one of our MCPR experiments, we studied the effectiveness of our signal manipulation attack and therefore attached one node to the relay hardware through a coaxial cable. While this configuration does not represent a realistic attack scenario, it simplified evaluation and allowed systematic evaluation of the intended attack. In all of our other experiments, the attacker picked up the victim signals wirelessly in ordinary  wireless environments.

\subsection{Attacker Capabilities and Improvements}
We assess the complexity of the attack setup to be moderate. As it is external to the legitimate parties and only uses off-the-shelf low-cost RF components, it can be realistically implemented by others. Although the setup is already capable of communication range extension and adaptive signal manipulation, further improvements can be made. Complementing the current mixed-signal processing with a digital receiver would allow to roughly track the nodes' protocol state. Eliminating signal direction finding based on power detection, RF circulators could be used to separate RF signals depending on their direction (we further discuss the use of circulators in Appendix~\ref{sec:appdx_circulator}). Currently, the attacker relies on a cabled connection between the two relay stations. Clearly, this could be replaced with a wireless link, although requiring additional engineering efforts. A challenge would be the isolation between the receiving and re-transmitting antennas of each relay station. This could be achieved through frequency conversion~\cite{Francillon.2010}, full-duplex radios~\cite{Bharadia.2014}, or directional antennas pointing away from each other at a distance~\cite{Lee2012IsolationEB}. An approach based on the latter will be part of a future publication. If bidirectional amplification is required at both relay stations, another aspect would be the wireless synchronization between the relay stations. For this, the primary power detector could trigger a radio transmitter (at another frequency) to which the secondary station listens to toggle the communication direction upon detection.

\subsection{Attack Detection and Mitigation}
A physical-layer relay attack ideally does not affect the application data but only alters physical quantities. Thus, a detection mechanism should likewise be deployed on the physical layer. This alone poses a hurdle to many devices already in the field as physical-layer data typically is not reported to the application or is not measured at all. Moreover, low-level signal processing is mostly implemented in a performance-optimized but less flexible manner, \eg, in hardware, making it difficult to retrofit detection mechanisms. 

\Paragraph{MCPR} %
As our results show, MCPR can be used to tackle range-extension-only attacks. However, we also showed that arbitrary distance manipulation is possible with simple off-the-shelf RF components. Since linear frequency sweeping is not mandatory for MCPR, an ad hoc security improvement to the currently available implementations would be a secure random frequency hopping\footnote{Likely at the cost of increased measurement time due to longer PLL settling times.}. Although this would render attacks more difficult, it would not be a sound security measure, since frequency can easily be measured, for example by using frequency counting or a phase-frequency detector. Specific to our phase shift scheduling mechanism, the legitimate nodes could introduce agreed-upon fake transmissions to make the attacker falsely proceed to the next phase shifter setting. To impede the attacker's power detection, the victim device could transmit with very low power. However, the attacker then could utilize an improved power detector or position closer to the victim. 

\Paragraph{ToF Measurement} Another possibility would be the addition of a direct ToF measurement by means of a challenge-response based distance bounding protocol. Fortunately, proprietary implementations have been reported for various BLE transceivers~\cite{ti.rtls, giovanelliRSSITimeofflightBluetooth2018, AbidinSecureAccuratePractical_2021}. However, due to the limited signal bandwidth of Bluetooth, it is unlikely to achieve measurement accuracy as high as with MCPR. %
Further, the rather slow-transient Gaussian pulse shape employed for Bluetooth could facilitate early symbol detection attacks. Still, predicting cryptographically secured challenges early certainly poses higher hurdles to attackers than manipulating MCPR measurements where the measurement signals do not carry meaningful information. Thus, we believe this could be a good starting point to impose higher attack complexities. This likewise is acknowledged by recent work of Abidin~\etal~\cite{AbidinSecureAccuratePractical_2021} who proposed a ranging system that combines MCPR with ToF to implement a distance bounding protocol specifically geared towards BLE applications.

\Paragraph{Secondary observations} Specific to PaaK-based PKE applications, smartphones should not constantly emit unlock commands to mitigate the risk of relay attacks (as we experienced for the smart lock). That is, the smartphone should first apply plausibility checks, \eg, based on sensor readings. Secondary tracking and surveillance systems of connected cars could report successful attacks. For instance, stolen cars may report location data and camera footage over a cellular connection. However, this could be bypassed, \eg, using wireless jamming.

\subsection{Future Work}
In this work, we presented an analog physical-layer relay attack capable of RF range extension for Bluetooth communication and adaptive signal manipulation. Based on our results, we outline possible directions for future work.

Security discussions on relay attacks are often based on vague assumptions on attacker capabilities, making it difficult to realistically assess the risk of attacks. Future work should define an attack taxonomy to properly categorize attacks and countermeasures.

While the physical layer provides the basis for proximity verification, an actual implementation lives within a possibly complex protocol. Thus, in conjunction with physical-layer analysis, proximity verification should be examined for potential weaknesses on the protocol-level.

Our current attack implementation serves as a proof-of-concept. Naturally, the hardware setup leaves room for improvements and we are currently in the process of investigating wireless relay links. %

\section{Conclusion}
\label{sec:conclusion}
In this paper, we introduced a novel concept to accomplish real-world relay attacks on \SI{2.4}{\GHz} TDD communications such as Bluetooth. Using a setup built from off-the-shelf RF components, we carried out successful relay attacks on a car and a smart lock. Our results highlight the need for a secure proximity verification which is currently lacking for Bluetooth. Therefore and in view of our attacker setup, we investigated the security of MCPR which recently found deployment aside of Bluetooth. We demonstrated the first practical on-the-fly phase manipulation attack on MCPR while simultaneously enhancing the communication range. Based on our findings, we suggest to implement MCPR with a mandatory frequency hopping to impede attacks. Finally, we hope that our work will raise awareness for relay attacks against Bluetooth-based proximity applications to accelerate the deployment of countermeasures.

\section*{Acknowledgements}
We thank our shepherd Aanjhan Ranganathan and the anonymous owner of the Tesla Model 3. This work was funded by the Deutsche Forschungsgemeinschaft~(DFG, German Research Foundation) under Germany’s Excellence Strategy - EXC 2092 CaSa - 390781972.

\bibliographystyle{ACM-Reference-Format}
\bibliography{refs}

\appendix
\section{Relay Hardware Details}
\label{sec:appendix}

\begin{figure}[h]
\centering
\hspace*{\fill}%
\subfloat[]{{%
        \includegraphics[width=0.48\columnwidth]{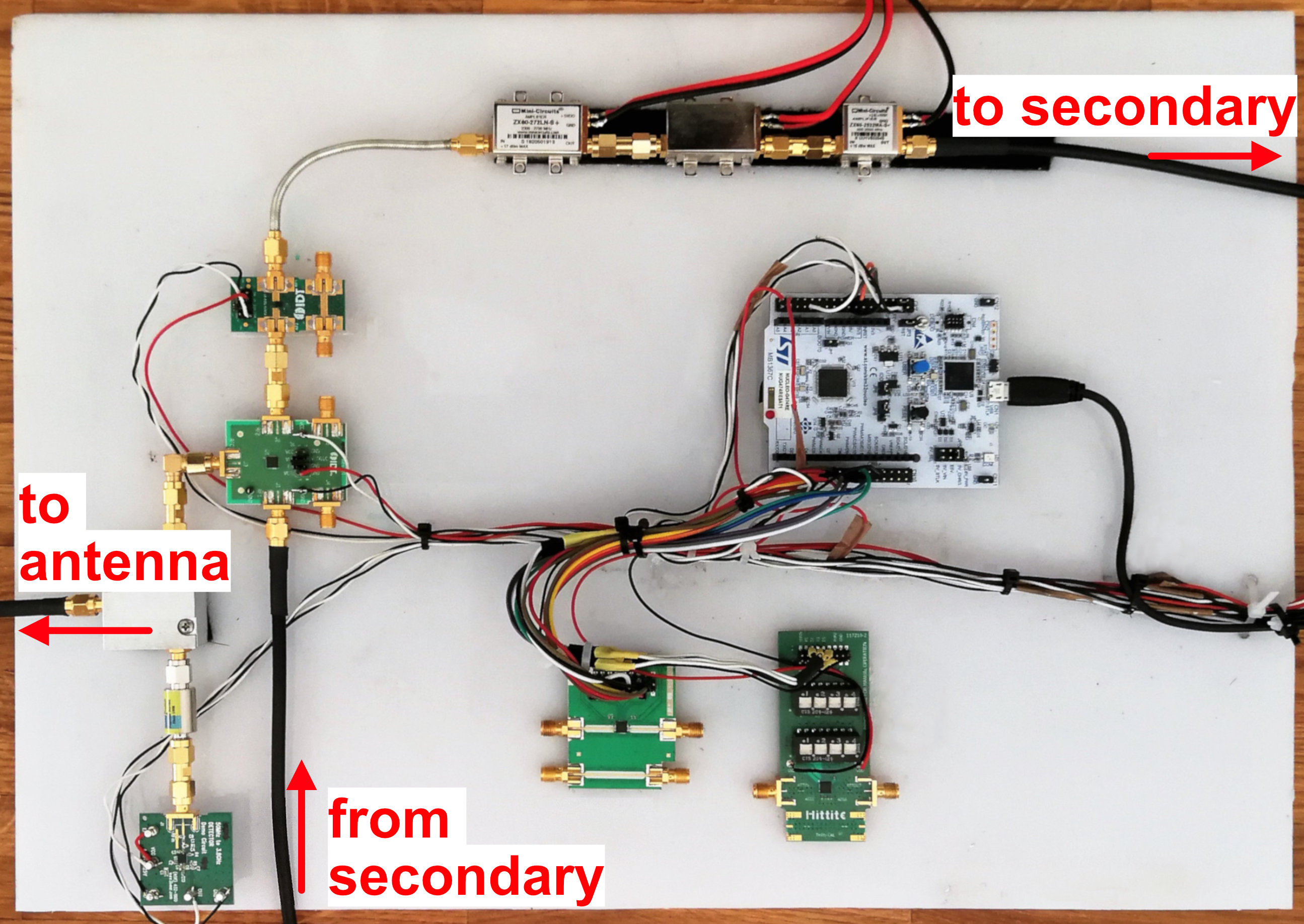}
        }}
\hfill
\subfloat[]{{%
        \includegraphics[width=0.48\columnwidth]{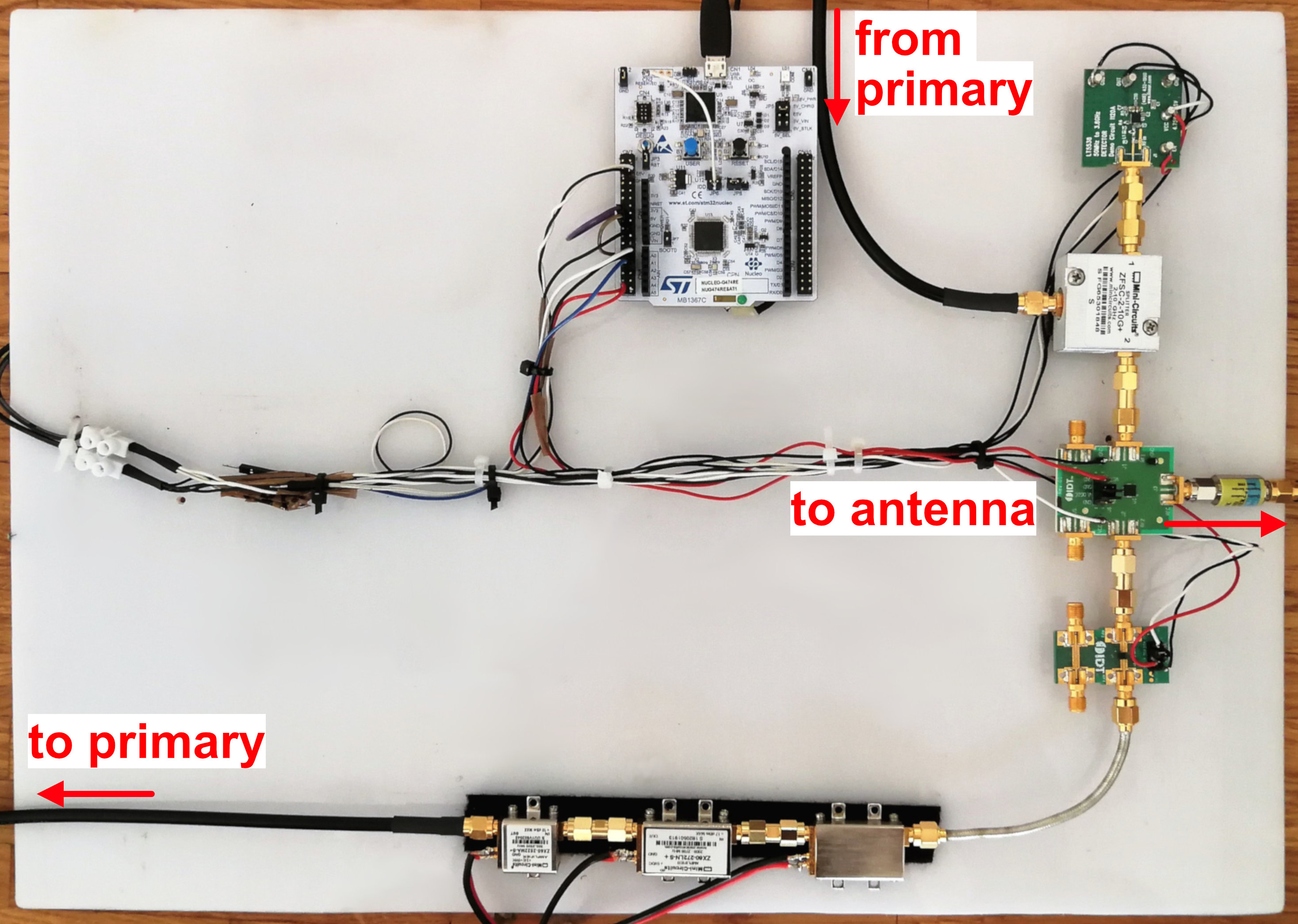}}}
\hspace*{\fill}%

\caption{Our experimental relay setup consists of the (a) primary and (b) secondary relay stations. At the primary station, a phase shifter and a step attenuator used in Section~\ref{sec:attack_evaluation} can be seen.}
\label{fig:analog_relay}
\end{figure}

\begin{table}[h]
\footnotesize\caption{Relay hardware components. The lower part lists components required to conduct the outlined signal manipulation attacks.}
\label{tab:relay_part_list}
\centering
\begin{tabular}{@{}rcl@{}}
\toprule
\textbf{Hardware Component} & \textbf{Quantity} & \textbf{Purpose}\\
\midrule
Mini-Circuits ZX60-272LN-S+ & 2                 & \tline[l]{Low-Noise Amplifier~(LNA),\\first amplifier stage}\\
Mini-Circuits ZX60-2534MA+  & 2                 & Amplifier\\
Mini-Circuits ZX60-2522MA+  & 2                 & Amplifier\\
Renesas F2932               & 2                 & \tline[l]{SPDT RF switch,\\signal direction selection}\\
Renesas F2910               & 2                 & \tline[l]{SP1T RF switch,\\isolation enhancement}\\
Mini-Circuits ZFSC-2-10G+   & 2                 & \tline[l]{RF power splitter,\\tap signal for power detector}\\
Analog Devices LT5538       & 2                 & \tline[l]{RF power detector,\\signal detection of $\mathsf{A}$}\\
Crystek CBPFS-2441          & 1                 & \tline[l]{Bandpass filter,\\reject noise and outband interference}\\ %
Siretta LLC200A 15m         & 8                 & \tline[l]{Coaxial cable,\\connect relay stations}\\ %
Interline PANEL 17          & 2                 & \tline[l]{Directional antenna,\\pick up victim signals}\\
\midrule
ST STM32G474RE              & 1                 & \tline[l]{Microcontroller,\\comparator and relay control}\\
M/A-COM MAPS-010164         & 1                 & \tline[l]{Phase shifter,\\MCPR distance manipulation}\\
Analog Devices HMC624A      & 1                 & \tline[l]{Step attenuator,\\reciprocity manipulation}\\
\bottomrule 
\end{tabular}
\end{table}

\begin{figure}[h]
\centering
\includegraphics[width=\linewidth]{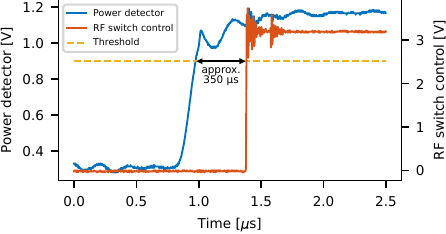}%
\caption{Output voltage of the power detector at the primary relay station, detection threshold voltage (\SI{0.9}{\V}, corresponding to an RF input power of approx.~\SI{-40}{dBm}~\cite{AnalogDevices.}), and RF switch control signal during the start of a Bluetooth transmission, indicating a reaction time of approx.~\SI{350}{\us} after threshold crossing.}
 \label{fig:detection_speed}
\end{figure}

\begin{figure}[h]
\centering
\includegraphics[width=\linewidth]{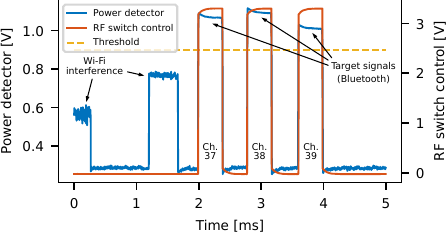}%
\caption{Power detector output voltage and RF switch control signal while receiving Bluetooth advertisement packets (consecutively on channels $37$, $38$, $39$) of the smart lock. Nearby Wi-Fi interference is received weaker and does not trigger detection.}
 \label{fig:advertising_interference}
\end{figure}

In the following, we provide additional details on our proof-of-concept attacker implementation. Figure~\ref{fig:analog_relay} depicts our experimental realization of the primary and secondary relay stations, assembled using off-the-shelf RF building blocks. An architectural block diagram of the implemented setup is shown in Fig.~\ref{fig:relay_block_only_PD}. For simplicity, we only show two amplifiers per direction in the block diagram. In fact, we used a cascade of three RF amplifiers, delivering a total gain of \SI{75}{\decibel} for each direction. A detailed list of all parts and their purpose to implement our attacker setup can be found in Table~\ref{tab:relay_part_list}. The upper part of the table lists the parts used for the relay configuration used in Section~\ref{sec:attacks}. For the experiments in Section~\ref{sec:attack_evaluation}, we additionally added the parts listed in the lower part of the table as shown in Fig.~\ref{fig:relay_block_with_phase}.

We selected the components listed in Table~\ref{tab:relay_part_list} primarily for their operating frequency to match the~\SI{2.4}{\GHz} frequency band. A fast response time is desired for the reactive switching of the relay to avoid cutting off initial parts of a transmission from~$\mathsf{A}$. Therefore, we specifically chose components having rather fast response times being much shorter than the~\SI{1}{\us} and~\SI{0.5}{\us} symbol duration of transmissions of BLE at~\SI{1}{Mbps} and~\SI{2}{Mbps}. The RF response time is governed by the power detector, the comparator which converts the power detector output into a logic signal, and the RF switches. The relay's response upon a beginning transmission of~$\mathsf{A}$ can be seen in Fig.~\ref{fig:detection_speed}. Reactive switching is accomplished within approx.~\SI{0.35}{\us} after signals from~$\mathsf{A}$ cross the power detection threshold at the primary relay station. Please note that this is approx.~0.795\% and 0.016\% of the shortest (\SI{44}{\us}) and longest (\SI{2128}{\us}) possible BLE packets, respectively~\cite{BluetoothSIG.January2019}. The relay directions are flipped back after A stops to transmit, without any limitation of the maximum transmission duration.
\balance
The power detector is capable of detecting signals from~\SIrange{-75}{10}{dBm}. Considering interference from other radio traffic, in a typical office environment, we found input powers of~\SI{-45}{dBm} to be sufficient to distinguish between targeted and other signals. Fig.~\ref{fig:advertising_interference} shows an example trace captured from the relay's primary power detector while the relay receives BLE advertisement packets from the smart lock (at \SI{1}{\m} distance). It can be seen that the high received signal power triggers the detection. In contrast, interfering signals from a nearby \mbox{Wi-Fi} access point (\SI{0}{} to \SI{2}{\ms} in the plot) are not as strong as the target signals and do not trigger the detection. 

For adaptive signal manipulation, \eg, MCPR distance manipulation, we used a microcontroller to process and react to the comparator output signals. The microcontroller accordingly controlled a phase shifter with $6$~bit phase resolution over~\SI{360}{\degree}~\cite{MACOM.}.

Another important aspect of the implementation is the isolation between the relay's transmission paths. Isolation is critical since a portion of the amplifier output will reach the input of the other direction's amplifiers. When the overall amplification is too large in regard of the finite path isolation, this again creates a feedback loop, causing instability of the overall system. In our relay implementation, the isolation between the paths is dominated by the electronically controlled solid state RF switches. Here, we initially were facing the RF amplifiers to slightly oscillate. We tackled the issue by adding a pair of F2910 SP1T switches in front of the LNA inputs to increase the total isolation. For the sake of simplicity, we do not indicate these switches in Fig.~\ref{fig:relay_block_only_PD} as these are always controlled in conjunction with the main SPDT switches.  

\section{MCPR Power Detector Output}
\label{sec:appdx_ranging_detector_output}

\begin{figure}[h]
\centering
\includegraphics[width=0.96\linewidth]{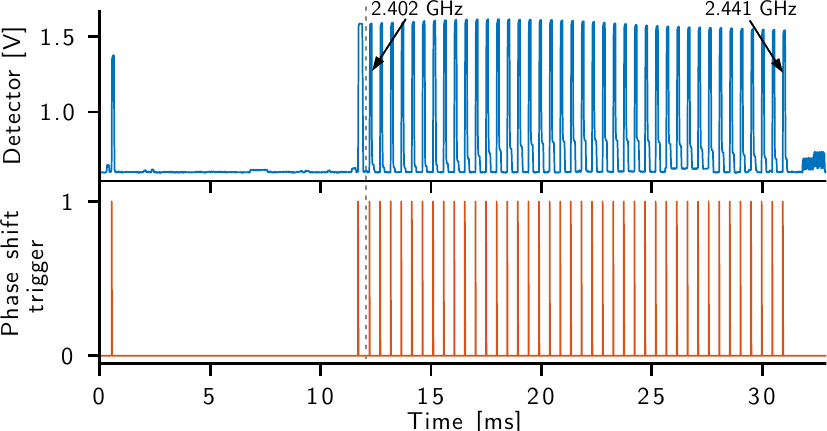}
\caption{Top: Raw power detector output signals, indicating when node~$\mathsf{A}$ transmits. Bottom: Detection times of the attacker to proceed to the next phase shift setting. From \SI{12}{\ms} on (dashed line), the MCPR tone exchange comprising of $40$~sweeped carriers is clearly visible.}
\label{fig:relay_ranging_attack_synchronization_update}
\end{figure}

Fig.~\ref{fig:relay_ranging_attack_synchronization_update}~(top) shows the power detector output before and during a proprietary MCPR procedure between two BLE transceivers~\cite{Dialog.ranging}. We clearly observe the effect of the sweeped carrier measurement comprising of $N=40$~successive tone transmissions. The rising edges of the power detector output~(bottom) correspond to the times $t$ at which we apply the phase shift $\phi_t$.

\section{Using Circulators}
\label{sec:appdx_circulator}
The TDD operation of the legitimate nodes is an important aspect for the implementation of our attacker setup. Although the attacker cannot predict when each node will transmit, the TDD implies that transmissions do not occur simultaneously. Importantly, this allows us to separately handle signal directions by means of adaptive switching based on power detection. As an alternative, it would be possible to replace the switches through circulators. A circulator is a passive three-port RF circuit, designed to separate signal directions. It is often used to transmit and receive over the same antenna simultaneously, which is desired for, \eg, radar or full-duplex radios. In particular, the circulator isolates the receiver from the transmitter and thereby mitigates interference. In the context of our attack, circulators would allow us to omit the power detection while applying bidirectional amplification. However, as we have outlined in Section~\ref{sec:attack_evaluation}, the power detector is essential to apply targeted signal manipulation. In comparison to RF switches, circulators typically are rather limited in terms of isolation~\cite{Bharadia_fullduplex}. This, however, limits the relay's amplification gain (and thus the maximum relaying distance) that can be applied without running into instability issues.

\end{document}